\documentclass[journal=jacsat,manuscript=article]{achemso}

\usepackage[version=3]{mhchem} % Formula subscripts using \ce{}

\usepackage{xcolor}

\usepackage{bm}
\usepackage{amssymb}
\usepackage{braket}
\usepackage{float}
\newcommand{\vect}[1]{\bm{\mathit{#1}}}

\author{Torsha Moitra}
\affiliation{Department of Chemistry, Indian Institute of Technology Bhilai, Durg 491002, Chhattisgarh, India.}
\email{torsha@iitbhilai.ac.in}

\title[]
  {Real-Time Time-Dependent Density Functional Theory for Pump-Probe Spectroscopies}

\begin{document}


\begin{abstract}
The last decade has witnessed a rapid advancement in laser technology, enabling the direct monitoring and control of electronic motion on its natural attosecond to sub-femtosecond timescales. Ultrafast processes are conventionally studied using pump-probe spectroscopic techniques, where a pump pulse drives the molecule out of equilibrium and a time-delayed probe pulse records the response of the coherent non-stationary state. Since, these processes are non-linear and non-perturbative in nature, real-time formalisms provide a suitable theoretical framework for studying ultrafast light-induced dynamics. In addition, relativistic effects can play an important role in such simulations, either because the external field lies in the XUV to soft-X-ray region targeting core-level excitations, or because the molecular system contains heavy elements. In this chapter, we provide an overview of recent developments in real-time time-dependent density functional theory for simulating pump-probe spectroscopies (namely, transient absorption and transient electronic circular dichroism) at both non-relativistic and relativistic Hamiltonian levels. In order to further interpret these spectroscopic signals, we analyze several spectroscopically relevant time-dependent sub-observables, such as induced electronic densities and induced dipole moments as well as analytical formulations of generalized non-equilibrium response functions. We provide examples to show that the framework can be used to investigate and design new light-induced phenomena that emerge only in the attosecond regime.
\end{abstract}

\section{Introduction}
Advances in laser technology has enabled the generation of laser pulses with unprecedented spatiotemporal resolution, making it possible to probe matter with an unprecedented level of temporal precision of attosecond scales ($10^{-18}$ s).~\cite{Corkum2007,Nisoli2017,li2020real,SverdrupOfstad2023}
The natural timescale of nuclear motion lies in the femtosecond regime, while the motion of electrons, being much lighter and faster, is an attosecond phenomenon. Hence, in this scenario the nuclei can be considered as frozen, and the entire dynamics is governed by electronic motions. 

The conventional tool for studying these ultrafast processes is pump-probe spectroscopy.~\cite{Kretschmar2024,LaraAstiaso2018,Guo2024,Koll2022} In this approach, a pump laser pulse drives the system out of equilibrium into a coherent superposition of electronic states or an electronic wavepacket. A second, time-delayed probe pulse then captures the evolving electronic structure by recording the time-dependent spectroscopic signal. Pump-probe spectroscopy offers several additional degrees of freedom compared to conventional single-pulse experiments, such as, pump pulse parameters (like polarization, carrier frequency, pulse shape, pulse strength) as well as the time-delay between the pump and probe pulses, all of which are imprinted onto the spectroscopic signal. However, the pump pulse needs to be sufficiently strong in order to depopulate the stationary ground state, and hence it operates in the non-perturbative regime. 
Capturing all these requires a theoretical framework that propagates the full electronic wavepacket or density in real-time, explicitly including both the pump and probe pulse parameters without expansion in any perturbation parameter.

Real-time propagation methods are the natural theoretical choice for this problem.~\cite{Kadek2024,li2020real}
In the real-time formalism, the time-dependent Schr\"odinger or Dirac equation is propagated in time, with the time-dependent external field entering the Hamiltonian expression. This provides a fully non-perturbative, time-domain description of the electronic response that is capable of accounting for arbitrary pulse features. An additional advantage is that the calculated time-dependent observables, such as induced dipole moments, charge densities, and current densities, provide a direct dynamical picture of the underlying electronic mechanism.

The development of real-time electronic structure methods has a long history. The theoretical roots lie in the time-dependent Hartree-Fock equations developed in the late 1980s.~\cite{Yamamura1987} A major step forward came with the multi-configuration time-dependent Hartree (MCTDH) method, developed by Cederbaum and co-workers, which propagates multi-dimensional nuclear or electronic wavepackets using a variationally optimized set of time-dependent basis functions and remains one of the most accurate and widely used methods for quantum electron-nuclear dynamics.~\cite{Meyer1990}
The real-time formulation of time-dependent density functional theory (RT-TDDFT) was introduced by Yabana and Bertsch, who demonstrated that the real-time propagation of the Kohn--Sham equation provides an efficient and broadly applicable route to optical response properties of molecules and solids.~\cite{yabana1996,Yabana2006}
Their original implementation used real-space grid to represent the Kohn--Sham orbitals, and was later extended to periodic-systems. The real-space grid formulation is particularly well suited for strong-field and non-perturbative dynamics, as it avoids basis-set incompleteness errors and treats all regions of space on equal footing.~\cite{DeGiovannini2013,castro2006,Andrade2012}
In parallel, the molecular quantum chemistry community developed RT-TDDFT implementations based on localized Gaussian type atomic orbital basis sets, benefiting from compact basis sets, systematic convergence, and easy interfacing with existing modules.~\cite{lopata2011modeling,Lopata2012,yang2022intruder,provorse2016electron} More recently, RT-TDDFT has also been formulated in numerical atom-centered orbital basis  as well as plane-wave basis enabling large-scale simulations of periodic systems.~\cite{hekele2021all,Qian2006,Walker2007,Walter2008,Schleife2012} 
Together, these complementary basis set strategies 
%- real-space grids, localized Gaussina or numerical atom-centered orbitals, plane waves and hybrid representations- 
give RT-TDDFT a broad applicability ranging from molecular systems to periodic systems. 
The development in real-time methods was also mirrored by the wavefunction based electronic structure community, leading to the formulation of real-time multi-configurational methods, including the time-dependent complete-active-space self-consistent field (TD-CASSCF)~\cite{Sato2013,Miyagi2013,Miyagi2014,Sato2015}, real-time configuration interaction (RT-CI)~\cite{CI1,CI2,CI3,CI6,CI7}, real-time coupled cluster (RT-CC)~\cite{CC0,CC1,CC2,CC3,CC4,CC8,CC12,CC11} methods. Although wavefunction-based real-time methods are available and formally appealing, their computational cost limits their routine application to larger systems. Time-dependent density functional theory offers a practical compromise between computational efficiency and accuracy and serves as a work-horse quantum chemistry method.
 
 Further, the demands of heavy-element chemistry and core-level spectroscopy necessitated the development of relativistic real-time methods. Repisky and coworkers reported the first implementation of RT-TDDFT at the fully relativistic four-component Dirac--Coulomb level, enabling variationally correct treatment of scalar relativistic effects and spin-orbit coupling within the real-time framework.~\cite{repisky2015excitation,Kadek2015} This was followed by the development of two-component relativistic real-time methods based on the exact two-component (X2C) Hamiltonians, which reduced computational cost significantly.~\cite{Konecny2016,moitra2023accurate,Kasper2018Modeling} Relativistic real-time methods based on coupled cluster theory have also been proposed, extending the hierarchy of relativistic methods for real-time dynamics.~\cite{CC6} 
 Together, this family of methods spans a broad range of accuracy and computational cost, from mean-field RT-TDDFT to correlated RT-CC and from non-relativistic to fully four-component relativistic treatments.

This chapter provides a comprehensive overview of RT-TDDFT for the simulation of pump-probe spectroscopies at both non-relativistic and relativistic Hamiltonian levels. In the following sections, the readers are introduced to the working equation of RT-TDDFT called Liouville-von Neumann equation of motion. Then details of the Hamiltonian or Fock representations, incorporating relativistic treatment, are described. Next, we list the computational steps involved in pump-probe spectroscopic simulations, including the incorporation of explicit pulse profiles. In order to analyze the spectroscopic observations we showcase two strategies: first, based on pseudo-observables that give visual qualitative understanding of the underlying mechanism and second, generalized non-equilibrium response theory which provides a mathematical justification for the unique signals associated with pump-probe spectroscopies. Finally, we present example applications of RT-TDDFT. 

\section{Liouville-von Neumann Equation of Motion}\label{sec:LvN}

The time evolution of a many-electron quantum system is governed by the time-dependent Schr\"odinger equation (TDSE),
\begin{align}
    i\hbar \frac{\partial}{\partial t} |\Psi(t)\rangle = \hat{H}(t) |\Psi(t)\rangle,
    \label{eq:tdse}
\end{align}
where $|\Psi(t)\rangle$ is the total wave function and $\hat{H}(t)$ is the time-dependent Hamiltonian describing both the molecular system as well as its interaction with the time-dependent external field.
In the context of attosecond and few-femtosecond (less than $10$ fs) timescales, the nuclear coordinates can be treated as frozen, and the problem reduces to the propagation of the \emph{electronic} wave function alone under the Born--Oppenheimer approximation. The TDSE thus becomes
\begin{align}
i\hbar \frac{\partial}{\partial t} |\Psi_{\mathrm{el}}(t)\rangle = \hat{H}_{\mathrm{el}}(t)|\Psi_{\mathrm{el}}(t)\rangle,
\end{align}
where \(|\Psi_{\mathrm{el}}(t)\rangle\) is the many-electron wave function and \(\hat{H}_{\mathrm{el}}(t)\) is the electronic Hamiltonian, comprising the electronic kinetic energy, electron-nuclear attraction (with fixed nuclear positions), electron-electron repulsion, and the time-dependent external field coupling.

Direct solution of the many-electron TDSE is computationally prohibitive for all but the smallest of systems. Time-dependent density functional theory (TDDFT) provides a formally exact and computationally feasible reformulation by replacing the many-body wave function with the time-dependent electron density $\rho(\bm{r},t)$ as the fundamental variable. 
The theoretical foundation of TDDFT rests on two key theorems: the Runge--Gross theorem,~\cite{runge1984density} which establishes a one-to-one correspondence between time-dependent external potentials and densities for a given initial state, and the van Leeuwen theorem,~\cite{vanLeeuwen1999mapping} which shows that the density evolution of an
interacting many-electron system can, in principle, be reproduced by an
auxiliary non-interacting Kohn--Sham (KS) system.

In the time-dependent KS scheme, the density is constructed from a set of non-interacting single-particle orbitals $\{\phi_p(\bm{r},t)\}$ evolving under an effective time-dependent potential. The TDKS equation thus becomes 
\begin{align}
i\hbar \frac{\partial}{\partial t} |\phi_i(t)\rangle
=
\hat{F}(t)|\phi_i(t)\rangle,
\end{align}
where $\hat{F}(t)$ is the time-dependent KS Fock operator. 
Expanding the time-dependent spin-orbitals in a finite set of orthonormal basis functions \(\{X_\mu(\bm{r})\}\),
\begin{align}
\phi_i(\vect{r},t)=\sum_{\mu} X_\mu(\vect{r}) C_{\mu i}(t),
\label{eq:KSorb-expansion}
\end{align}
the one-electron reduced density matrix is written as
\begin{align}
D_{\mu\nu}(t)=\sum_{i}^{\mathrm{occ}} C_{\mu i}(t) C_{\nu i}^{*}(t),
\end{align}
or, in matrix notation,
\begin{align}
\mathbf{D}(t)=\mathbf{C}(t)\mathbf{C}^{\dagger}(t).
\label{eq:RDM-matrix}
\end{align}
The initial density matrix $\mathbf{D}(t=0)$ is obtained by solving the time-independent Kohn-Sham self-consistent field equations for the ground state of the system.
After simple mathematical manipulations, we arrive at the time-dependent equation-of-motion for the reduced density matrix, as
\begin{align}
i\hbar\frac{\partial \mathbf{D}(t)}{\partial t}
=
[\mathbf{F}(t),\mathbf{D}(t)],
\label{eq:LvNEOM}
\end{align}
where \([\mathbf{A},\mathbf{B}]=\mathbf{A}\mathbf{B}-\mathbf{B}\mathbf{A}\) denotes the commutator. For a single-determinant wave function, the entire time evolution is therefore encoded in the one-electron reduced density matrix. Eq.~\ref{eq:LvNEOM} is known as the Liouville-von Neumann (LvN) equation of motion.
 The advantage of this formulation is that the density matrix is invariant with respect to unitary rotations within the occupied orbital space and provides direct access to one-electron observables through simple trace expressions.

A key challenge in propagating the density matrix arises from the self-consistent nature of the problem, that is, the Fock operator at time $t$ depends on the density matrix at that same time, as evident from Eq.~\ref{eq:LvNEOM}. This necessitates careful construction of numerical propagators that preserve the essential physical properties of the density matrix, namely unitarity, time-reversal symmetry, and trace conservation, throughout the simulation. Over the years, several propagator schemes have been developed to meet these requirements.~\cite{castro2004propagators,Ye2024,crank1947practical,schelter2018accurate,pedersen2019symplectic} The modified midpoint unitary transformation (MMUT) provides a computationally efficient approach by evaluating the Fock matrix at the midpoint of each time step.~\cite{liang2011efficient,li2005mmut} Magnus expansions offer systematic improvements in accuracy through higher-order corrections, with the second-order variant being widely adopted for its balance between precision and cost.~\cite{blanes2006fourth,alvermann2011high,gomez2020propagators,repisky2015excitation} Chebyshev polynomial expansions represent an alternative strategy particularly suited for systems with large energy ranges.~\cite{williams2016accelerating} 

\section{Fock Matrix and Hamiltonian Representations}\label{sec:Fock}

The propagation of Liouville-von Neumann equation of motion (Eq.~\ref{eq:LvNEOM}) requires the construction of the time-dependent Fock matrix at each time step. In the orthonormal basis $\{X_\mu(\bm{r})\}$, it is written formally as
\begin{align}
F_{\mu\nu}(t)
=
h_{\mu\nu}
+
G_{\mu\nu}[\mathbf{D}(t)]
+
V^{\mathrm{xc}}_{\mu\nu}[\rho(t)]
+
V^{\mathrm{ext}}_{\mu\nu}(t),
\label{eq:Fock-general}
\end{align}
where $h_{\mu\nu}$ denotes the one-electron Hamiltonian matrix elements, $G_{\mu\nu}[\mathbf{D}(t)]$ collects the two-electron Coulomb and exchange contributions, $V^{\mathrm{xc}}_{\mu\nu}[\rho(t)]$ is the exchange--correlation (xc) contribution, and $V^{\mathrm{ext}}_{\mu\nu}(t)$ describes the coupling to the external time-dependent field. 
The Fock matrix is time-dependent both implicitly through the density matrix and explicitly through the time-dependent external perturbation. 
For pump--probe simulations, we introduce the interaction with the external perturbation within the electric-dipole approximation and length gauge as,
\begin{align}
    V^\text{ext}_{\mu\nu}(t)  = -\sum_{u\in x,y,z}P_{u,\mu\nu} \mathcal{E}_u(t)
    -\sum_{u\in x,y,z}P_{u,\mu\nu} \mathcal{F}_u(t)
\end{align}
where the pump and probe electric fields are represented as $\bm{\mathcal{E}}$ and $\bm{\mathcal{F}}$, respectively. The electric dipole moment operator matrix element is given by $P_{u,\mu\nu}$.
This compact form (Eq.~\ref{eq:Fock-general}) is common to non-relativistic (1c) , two-component (2c), and four-component (4c) real-time frameworks; the differences lie in how the one-electron Hamiltonian, two-electron terms, and density-dependent quantities are represented.

At the fully relativistic level, the underlying one-electron operator is the Dirac Hamiltonian, and the resulting four-component (4c) Dirac--Coulomb Fock matrix provides the most rigorous description of scalar-relativistic and spin--orbit effects within RT-TDDFT.
The main computational bottleneck of the 4c formulation is the evaluation of the generalized two-electron integrals~\cite{ReSpect}
\begin{align}
    & G^\mathrm{4c}_{\mu\nu,\kappa\lambda} = \mathcal{I}^\mathrm{4c}_{\mu\nu,\kappa\lambda} - \zeta \mathcal{I}^\mathrm{4c}_{\mu\lambda,\kappa\nu}
    \quad ; \quad
\mathcal{I}^\mathrm{4c}_{\mu\nu,\kappa\lambda} = \iint \Omega^{\mathrm{4c}}_{0,\mu\nu} (\vect{r}_1) r_{12}^{-1} \Omega^{\mathrm{4c}}_{0,\kappa\lambda} (\vect{r}_2) d^3\vect{r}_1 d^3\vect{r}_2~,\\
    & \text{where},\quad 
    \Omega^{\mathrm{4c}}_{0,\mu\nu} (\vect{r}) = X^\dagger_\mu(\vect{r})X_\nu(\vect{r}) ~.
\end{align}
Since, the four-component basis ($X_\mu(\vect{r})$) contains both large- and small-component spinors,
 connected by the restricted kinetic balance (RKB) condition~\cite{Stanton1984}
\begin{align}
X^{\mathrm{S}}(\bm{r}) = \frac{\boldsymbol{\sigma} \cdot \mathbf{p}}{2mc} \, X^{\mathrm{L}}(\bm{r}),
\label{eq:RKB}
\end{align}
the resulting electron-repulsion integral machinery is substantially more complicated than in the non-relativistic case. As discussed in Ref.~\citenum{ReSpect}, even in a compact real-quaternion implementation, a single 4c electron-repulsion integral requires the simultaneous evaluation and processing of roughly 25 times more real scalar integrals than the corresponding 1c or simpler 2c case, making full 4c real-time propagation feasible but computationally demanding.

To reduce this cost while retaining the accuracy of the parent 4c description, the exact two-component (X2C) transformation can be applied to the relativistic equation of motion.~\cite{Heully1986,Kutzelnigg2005,Liu2007,Ilias2007} The X2C transformation is defined via a unitary decoupling matrix $\mathbf{U}$ that block-diagonalizes the 4c Fock and density matrix as
\begin{align}
\widetilde{F}^{\mathrm{2c}}_{\mu\nu}(t)
=
\left[\mathbf{U}^\dagger \mathbf{F}^{\mathrm{4c}}(t) \mathbf{U}\right]^{\mathrm{LL}}_{\mu\nu} 
\quad,\quad
\widetilde{D}^{\mathrm{2c}}_{\mu\nu}(t)
=
\left[\mathbf{U}^\dagger \mathbf{D}^{\mathrm{4c}}(t) \mathbf{U}\right]^{\mathrm{LL}}_{\mu\nu},
\end{align}
where LL denotes the large-component--large-component block. In principle, this yields a picture-change-transformed two-component Fock (and reduced density) matrix that is formally equivalent to the four-component one. Note that all transformed quantities are marked by tilde symbol ($~\widetilde{}$~). However, the fully transformed 2c expression still involves transformed two-electron and exchange-correlation quantities and may therefore be computationally even less attractive than the original 4c form. 
Stemming from this challenge, there are several variants of approximate Hamiltonian where the computationally demanding transformations are replaced by computationally efficient untransformed quantities.~\cite{AMFI,Goings2016,vanWllen2005,Peng2007,Filatov2013,Liu2018amf,Knecht2022} Out of the plethora of methods, 
we advocate the use of the 
 atomic mean-field exact two-component Hamiltonian (amfX2C) variant due to its accuracy in comparison to the gold-standard 4c method.~\cite{Knecht2022,moitra2023accurate}

In the amfX2C approach, the time-dependent Fock matrix is written as~\cite{moitra2023accurate}
\begin{align}
%\widetilde{F}^{\mathrm{2c}}_{\mu\nu}(t) 
%=
F^{\mathrm{amfX2C}}_{\mu\nu}
%[\mathcal{E}, \mathcal{F}]
(t)
&=
\widetilde{h}^{\mathrm{2c}}_{\mu\nu}
-
\sum_{u \in x,y,z}
\widetilde{P}^{\mathrm{2c}}_{u,\mu\nu}\,\mathcal{E}_u(t)
-
\sum_{u \in x,y,z}
\widetilde{P}^{\mathrm{2c}}_{u,\mu\nu}\,\mathcal{F}_u(t)
+
\Delta\widetilde{F}^{\mathrm{amfX2C}}_{\oplus,\mu\nu}
\nonumber\\
&+
\sum_{\kappa\lambda}
G^{\mathrm{2c}}_{\mu\nu,\kappa\lambda}\,\widetilde{D}^{\mathrm{2c}}_{\lambda\kappa}(t, \mathcal{E}, \mathcal{F})
+
\sum_{u \in 0\text{--}3}
\int
v^{xc}_{u}\!\left[\rho^{\mathrm{2c}}(\vect{r}, t, \mathcal{E}, \mathcal{F})\right]
\Omega^{\mathrm{2c}}_{u,\mu\nu}(\vect{r})\,d^3\bm{r},
\label{eq:Fock-amfX2C}
\end{align}
where $\widetilde{h}^{\mathrm{2c}}_{\mu\nu}$ is the X2C picture-change transformed one-electron Hamiltonian, $\widetilde{P}^{\mathrm{2c}}_{u,\mu\nu}$ are the transformed electric dipole moment matrices coupling to the pump ($\bm{\mathcal{E}}$) and probe ($\bm{\mathcal{F}}$) fields, and $\widetilde{D}^{\mathrm{2c}}(t, \bm{\mathcal{E}},\bm{\mathcal{F}})$ is the transformed density matrix. In contrast, the two-electron integral matrix $G^{\mathrm{2c}}_{\mu\nu,\kappa\lambda}$, the exchange--correlation potential $v^{xc}_{u}$, and the overlap distribution functions $\Omega^{\mathrm{2c}}_{u,\mu\nu}$ are evaluated using {untransformed} two-component basis functions. The term $\Delta\widetilde{F}^{\mathrm{amfX2C}}_{\oplus,\mu\nu}$ represents the picture-change correction, which accounts approximately for the missing two-electron and exchange-correlation picture-change effects through a superposition of atomic mean-field contributions
\begin{align}
\Delta\widetilde{\mathbf{F}}^{\mathrm{amfX2C}}_{\oplus}
=
\bigoplus_{K=1}^{M}
\left[\Delta
\widetilde{\mathbf{F}}^{\mathrm{2c,2e}}_{K}
+
\widetilde{\mathbf{F}}^{\mathrm{2c,xc}}_{K}
\right],
\end{align}
where the sum $K$, runs over all $M$ atoms in the molecule, and each atomic contribution is obtained from independent atomic SCF calculation in the AO basis of the $K$-th atom.

\section{Pump-Probe Spectra Simulation}\label{sec:pp_rt}

One of the main advantages of the real-time formulation is that the external perturbation can be defined directly in the time domain. This allows, in principle, the use of arbitrary pulse shapes, including finite-duration pulses, few-cycle pulses, chirped pulses, circularly polarized pulses, or impulsive perturbations. Recalling that in pump-probe simulations, the pump pulse is used to prepare a nonstationary electronic state, while the probe pulse is applied after a chosen time delay $\tau$ to capture the response of the evolving non-stationary wavepacket. The critical parameter is the spectral evolution as a function of pump-probe delay.

For the pump pulse, different functional forms may be used depending on the physical process of interest. A commonly used linearly polarized finite pulse is a $\sin$ carrier modulated by a $\cos^2$ envelope,
\begin{equation}
\boldsymbol{\mathcal{E}}(t)
=
\mathbf{n}\mathcal{E}(t)
=
\mathbf{n}\mathcal{E}_0
\cos^2\left(\frac{\pi}{T}(t-t_0)\right)
\sin(\omega_0 t+\phi),
\label{eq:pumpLP}
\end{equation}
where $\mathbf{n}$ defines the polarization direction, $\mathcal{E}_0$ is the field amplitude, $T$ controls the pulse duration, $t_0$ is the pulse center, $\omega_0$ is the carrier frequency, and $\phi$ is the carrier-envelope phase. 
For chiral or circularly polarized excitation, the pump can instead be written as a Gaussian-enveloped circularly polarized pulse,
\begin{equation}
\boldsymbol{\mathcal{E}}^{L/R}(t)
=
\begin{cases}
\mathcal{E}_0 \mathbf{e}^{L/R}
\exp\left[-\dfrac{(t-t_0)^2}{2\sigma^2}\right],
& t \leq T, \\
0,
& t > T,
\end{cases}
\label{ew:pumpCP}
\end{equation}
where $\mathcal{E}_0$ is the amplitude, $t_0$ is the pulse center, $\sigma$ is the temporal width of the Gaussian envelope, and $L/R$ denotes left- or right-circular polarization. The corresponding helicity vector may be written as
\begin{equation}
\mathbf{e}^{L/R}
=
\cos\left[\omega_0(t-t_0)\right]\mathbf{x}
\mp
\sin\left[\omega_0(t-t_0)\right]\mathbf{y},
\end{equation}
where the two signs correspond to opposite helicities of the circularly polarized field.
The probe pulse is conveniently chosen as an impulsive delta-like perturbation,
\begin{equation}
\boldsymbol{\mathcal{F}}(t)
=
\mathbf{m}\mathcal{F}(t)
=
\mathbf{m}\mathcal{F}_0\delta(t-(T+\tau)),
\label{eq:probe}
\end{equation}
where $\mathbf{m}$ is the probe polarization direction, $\mathcal{F}_0$ is the probe amplitude, $T$ is the end time of the pump pulse, and $\tau$ is the pump-probe delay. The use of a delta-function probe is particularly useful because it excites all frequencies simultaneously and therefore provides access to the broadband frequency-dependent response from a single real-time propagation.

The simulation of transient absorption spectroscopy (TAS)~\cite{moitra2023accurate} and transient electronic circular dichroism (TCD)~\cite{Moitra2025} spectra follows a common real-time protocol. The two cases differ mainly in the observable recorded after the probe perturbation and in the final spectral function constructed from the time-domain signal. A general computational protocol is as follows:

\begin{enumerate}
    \item Compute pump--probe (pp) as well as pump-only (p) induced electric or magnetic dipole moment $(u, v \in \{x, y, z\})$:
    \begin{align}
    \mu^{\mathrm{ind,pp/p}}_{uv}(t) = \mathrm{Tr}\left[\mathbf{P}_u \mathbf{D}^{\mathrm{pp/p}}_v(t)\right] - \mu^{\mathrm{static}}_u
    \qquad &\text{[TAS]}
    \\
    m^{\mathrm{ind,pp/p}}_{uv}(t) = \mathrm{Tr}\left[\mathbf{M}_u \mathbf{D}_v^{\mathrm{pp/p}}(t)\right] - m^{\mathrm{static}}_u
    \qquad &\text{[TCD]}
    \end{align}
    where $\mathbf{P}_u$ and $\mathbf{M}_u$ are the matrix representation of the electric and magnetic dipole operator, respectively.
     
    \item Compute the differential induced dipole moment:
    \begin{align}
    \Delta\mu^{\mathrm{TAS}}_{uv}(t) = \mu^{\mathrm{ind,pp}}_{uv}(t) - \mu^{\mathrm{ind,p}}_{uv}(t) 
    \qquad &[\mathrm{TAS}]
\\
    \Delta m^{\mathrm{TCD}}_{uv}(t) = m^{\mathrm{ind,pp}}_{uv}(t) - m^{\mathrm{ind,p}}_{uv}(t)
     \qquad &[\mathrm{TCD}].
    \end{align}
    This differential signal isolates the effect of the probe pulse acting on the nonstationary electronic wavepacket prepared by the pump, thereby removing the pump-only contributions. The time-domain differential dipole moment is subsequently Fourier transformed to the frequency domain, with appropriate damping to account for the finite simulation time.
    
    \item Evaluate spectral function as:~\cite{2009_Barron}
    \begin{align}
    &[\mathrm{TAS}]\qquad
    S^{\mathrm{TAS}}(\omega) = \frac{4\pi\omega}{3c}\,\Im ~\mathrm{Tr} \left[\boldsymbol{\alpha}^{\mathrm{TAS}}(\omega)\right], ~\mathrm{where,~} \alpha_{uv}^\mathrm{TAS} = \frac{\Delta \mu_{uv}^\mathrm{TAS}(\omega)}{\mathcal{F}_0} 
    \label{eq:TAS}
\\
 &[\mathrm{TCD}]\qquad
    \Delta\varepsilon^\mathrm{TCD}(\omega) = \frac{20 N_A\omega}{3\ln{(10)}c^2\epsilon_0} \Im ~\mathrm{Tr}[\boldsymbol{\beta}^\mathrm{TCD}(\omega)]
    ,~\mathrm{where,~} \beta_{uv}^\mathrm{TCD} = -i\frac{\Delta m_{uv}^\mathrm{TCD}(\omega)}{\mathcal{F}_0} 
    \label{eq:TCD}
    \end{align}
    Here, $\mathcal{F}_0$ is the probe field amplitude, $S^{\mathrm{TAS}}(\omega)$ represents the transient absorption cross section, and $\Delta\varepsilon(\omega)$ is the differential molar extinction coefficient expressing the circular dichroism signal. 
    In Eq.~\ref{eq:TAS} and \ref{eq:TCD}, $N_A$ is the Avogadro constant, $c$ is the speed of light in vacuum, $\epsilon_0$ is the vacuum permittivity.
    The trace over Cartesian components $u$ accounts for orientational averaging over randomly oriented molecules.
\end{enumerate}

\section{Pseudo-Observables: Induced Charge and Current Density}
In addition to experimentally measurable spectra, real-time TDDFT provides direct access to {pseudo-observables}, which are quantum mechanical quantities that encode ultrafast electron dynamics but are not directly measured. Amongst them the most insightful ones are the time-dependent induced charge density and current density, which reveal the spatial distribution and flow of electronic charge during ultrafast processes.

The induced charge and current densities are expressed as:~\cite{Moitra2025,Kadek2024}
\begin{align}
\rho^{\text{ind}}(\bm{r},t) &= \text{Tr}\left[ \mathbf{D}^{\text{ind}}(t) \, \boldsymbol{\Omega}(\bm{r}) \right] \\
j_k^{\text{ind}}(\bm{r},t) &= \text{Tr}\left[ \mathbf{D}^{\text{ind}}(t) \, \mathbf{J}_k(\bm{r}) \right], \quad k \in \{x, y, z\}
\end{align}
where $\mathbf{D}^{\text{ind}}(t) = \mathbf{D}(t) - \mathbf{D}_0$ is the induced one-electron reduced density matrix represented in the ground state molecular orbitals.
In the 1c nonrelativistic formalism, the charge and current density operator matrices are,
\begin{align}
\Omega^{1c}_{pq}(\bm{r}) &= \varphi_p^*(\bm{r}) \varphi_q(\bm{r}) \\
 J^{1c}_{k,pq}(\bm{r})
  &=
  -\frac{1}{2}
  \Big( \varphi_p^\dagger(\bm{r})\big\{p_{k}\varphi_q(\bm{r})\big\} 
     + \big\{p_{k}\varphi_p(\bm{r})\big\}^{\dagger} \varphi_q(\bm{r})   
  \Big).
\end{align}
where $\varphi_p(\bm{r})$ are scalar ground-state molecular orbitals and $p_k$ is the $k$-th component of the momentum operator.
While in the 4c relativistic formalism, the operator matrices are constructed
from the ground-state four-component molecular spinors
$
\psi_p(\mathbf r)=\sum_{\mu}X_\mu(\mathbf r)C^0_{\mu p},
$
where ${C}^0_{\mu p}$ is the ground-state 4c molecular spinor coefficient matrix elements and $X_\mu(\mathbf r)$ are the basis spinors containing large and small components. The charge,
spin-density, and current-density operator matrices in the ground-state molecular
spinor basis are~\cite{Kadek2024}
\begin{align}
\Omega^{4c}_{u,pq}(\mathbf r)
&=
\psi_p^\dagger(\mathbf r)\Sigma_u\psi_q(\mathbf r),
\qquad u=0,1,2,3,
\label{eq:omega4c_mo}
\\
J^{4c}_{k,pq}(\mathbf r)
&=
-c\,\psi_p^\dagger(\mathbf r)\alpha_k\psi_q(\mathbf r),
\qquad k\in\{x,y,z\}.
\label{eq:j4c_mo}
\end{align}
Here, $\Sigma_0=\mathbb{I}_4$ gives the scalar charge-density overlap distribution,
while
$
\Sigma_{1-3}%=
% \begin{pmatrix}
% \sigma_u & 0\\
% 0 & \sigma_u
% \end{pmatrix},
%  u=1,2,3,
$
gives the three spin-density components.
%, with \(\sigma_u\) being the Pauli
matrices. 
The matrices $\bm{\alpha}_k$ are the Dirac alpha matrices.

The visualization of electron charge density on a real-space grid is often used in the context of charge migration dynamics following sudden ionization process.~\cite{Bruner2017,Folorunso2021,Folorunso2023,Giri2023,Tremblay2022,Giri2022}
However, in the context of this chapter, we will present some results of charge and current density induced by an explicit pump pulse in section dedicated to examples.

\section{Non-Equilibrium Response Theory}\label{sec:NEQ}

Time-dependent density functional theory for stationary-state problems is conventionally formulated within the frequency-domain response theory framework, which provides a computationally efficient approach to computing excitation energies and spectroscopic properties of systems in their ground or excited eigenstates. However, this equilibrium response formalism is not directly applicable to pump-probe spectroscopies, where the system is first driven into a coherent superposition of states by the pump pulse and subsequently perturbed by a time-delayed probe. The transient nonstationary electronic wavepacket prepared by the pump evolves under field-free conditions and exhibits quantum coherences absent in any single eigenstate, leading to spectral features, such as time-delay-dependent oscillations and negative absorption peaks, that cannot be captured by conventional linear response theory. Nevertheless, the formal extension to a generalized nonequilibrium response theory formalism is rather straightforward and provides valuable interpretive insight into the origin of these unique pump-probe signatures. Although the real-time propagation approach described above is more practical for simulating time-resolved spectra, we detail the nonequilibrium response framework here for interpretive purposes, focusing on the role of quantum coherence and the extension to relativistic Hamiltonians with complex orbitals.~\cite{Perfetto2015}

Consider a nonstationary state $|\Psi[\mathcal{\bm{E}}]\rangle$ prepared by a pump pulse $\mathcal{\bm{E}}(t)$ lasting from time $0$ to $T$. After the pump ends, the system evolves under the static Hamiltonian $\hat{H}_0$ until a probe pulse $\mathcal{F}(t) = F_0\,\hat{\mathbf{m}}\,\delta(t - T - \tau)$ is applied at time $(T + \tau)$, where $\tau$ is the pump-probe delay. The time evolution for $t > T$ can be expressed using a sequence of unitary evolution operators. First, the system propagates freely under $\hat{H}_0$ from $t = T$ to $t = T + \tau$. Second, the $\delta$-type probe pulse instantaneously perturbs the wave function as $e^{-i\hat{Q}_v}$, where $\hat{Q}_v = -F_0\,\hat{\mathbf{P}}_v$ and $\hat{\mathbf{P}}_v$ is the electric dipole operator. Third, the system evolves freely from $t = T + \tau$ onward. The resulting wave function is~\cite{Tannor2007,Norman2018}
\begin{align}
|\Psi(t)\rangle = e^{-i\hat{H}_0(t - T - \tau)} e^{-i\hat{Q}_v} e^{-i\hat{H}_0\tau} |\Psi[\mathcal{\bm{E}}]\rangle.
\end{align}
We expand the pump-prepared state $|\Psi[\mathcal{\bm{E}}]\rangle$ in terms of the complete set of eigenstates $\{|\Phi_j\rangle\}$ of the field-free Hamiltonian $\hat{H}_0$ with corresponding eigenvalues $\varepsilon_j$,
\begin{align}
|\Psi[\mathcal{\bm{E}}]\rangle = \sum_j c_j[\mathcal{E}] |\Phi_j\rangle, \quad \text{where} \quad c_j[\mathcal{E}] = \langle \Phi_j | \Psi[\mathcal{\bm{E}}]\rangle.
\end{align}
The electric dipole response observable at time $t$ is given by $P_u[\mathcal{\bm{E}},\mathcal{\bm{F}}](t) = \langle \Psi(t) | \hat{P}_u | \Psi(t) \rangle$. Transforming this expectation value to the frequency domain and employing the resolution-of-identity relation $\sum_j |\Phi_j\rangle\langle \Phi_j| = \hat{1}$, we obtain the nonperturbative response,
\begin{align}
P_u[\mathcal{\bm{E}},\mathcal{\bm{F}}](\omega) = i\sum_{jkmn} c^*_j[\mathcal{\bm{E}}] c_n[\mathcal{\bm{E}}] e^{i\omega_{jn}\tau}  
\frac{(e^{iQ_v})_{jk}P_{u,km}(e^{-iQ_v})_{mn}}{\omega_{km} + \omega + i\Gamma},
\end{align}
where $\omega_{jk} = \varepsilon_j - \varepsilon_k$ denotes the energy difference between eigenstates and a phenomenological damping parameter $\Gamma$ has been introduced to ensure convergence of the Fourier transform for finite-time simulations.

To extract the leading-order response in the probe field strength, we expand the probe-induced evolution operators as $e^{- i\hat{Q}_v} \equiv e^{i\mathcal{F}_0P_v} = \mathbb{I} + i\mathcal{F}_0P_v + \mathcal{O}(\mathcal{F}_0^2)$. After simple mathematical rearrangements we obtain the first terms as,
\begin{align}
P^{(0)}_u[\mathcal{\bm{E}},\mathcal{\bm{F}}] (\omega) &=  i\sum_{jk}
c_j(\mathcal{\bm{E}})
c_k^*(\mathcal{\bm{E}})
e^{-\omega_{jk}\tau}
\frac{P_{u,kj}}{\omega-\omega_{jk}+i\Gamma} 
\nonumber \\
\frac{P^{(1)_u[\mathcal{\bm{E}},\mathcal{\bm{F}}] (\omega)}}{\mathcal{F}_0} &= \sum_{jkm} 
c_j(\mathcal{\bm{E}})
c_k^*(\mathcal{\bm{E}})
e^{-\omega_{jk}\tau}
\frac{P_{v,km}P_{u,mj}}{\omega-\omega_{jm}+i\Gamma}
+ \mathrm{c.c.} (\omega \to -\omega)
\end{align}
where the notation c.c.$(\omega \to -\omega)$ indicates the addition of the complex conjugate with simultaneous replacement of $\omega$ by $-\omega$. This result differs fundamentally from the equilibrium case, where the system begins in a single eigenstate $|\Phi_0\rangle$ (i.e., $c_j[\mathcal{\bm{E}}] = \delta_{j0}$), and coherences between different eigenstates are absent.

The frequency-dependent electric dipole-electric dipole response function is defined as $\chi_{\hat{P}_u\hat{P}_v}(\omega, T+\tau) = P_u^{(1)}[\mathcal{\bm{E}},\mathcal{\bm{F}}](\omega)/\mathcal{F}_0$, whose imaginary part determines the absorption cross section. To analyze the structure of this response function in the presence of spin-orbit coupling, we write the complex-valued expansion coefficients in polar form as $c_j = |c_j| e^{i\varphi_j}$, and introduce the time-delay-dependent phase
$\Theta_{jk}(\tau) = \varphi_j - \varphi_k - \omega_{jk}\tau$.
Additionally, using the relation,
$\frac{1}{\omega - \omega_{jm} + i\Gamma} = \mathcal{R}(\omega - \omega_{jm}) - i\mathcal{L}(\omega - \omega_{jm}),$
where $\mathcal{L}(\omega)$ is a Lorentzian lineshape function and $\mathcal{R}(\omega)$ is the associated Rayleigh (dispersive) component. Upon extracting the imaginary part, we arrive at
\begin{align}
\Im\chi_{\hat{P}_u\hat{P}_v}[\mathcal{\bm{E}}](\omega, T+\tau)] = 
\sum_{jkm} |&c_j[\mathcal{\bm{E}}]c_k[\mathcal{\bm{E}}]| \Re({P}_{v,km} {P}_{u,mj} ) \nonumber \\
\times \Big[
&\sin(\Theta_{jk}(\tau)\mathcal{R}(\omega-\omega_{jm})
-\cos(\Theta_{jk}(\tau)\mathcal{L}(\omega-\omega_{jm})
\nonumber \\ 
+&\sin(\Theta_{jk}(\tau)\mathcal{R}(\omega+\omega_{jm})
+\cos(\Theta_{jk}(\tau)\mathcal{L}(\omega+\omega_{jm})\Big]
\nonumber\\
+\sum_{jkm} |&c_j[\mathcal{\bm{E}}]c_k[\mathcal{\bm{E}}]| \Im({P}_{v,km} {P}_{u,mj} ) \nonumber \\
\times \Big[
&\cos(\Theta_{jk}(\tau)\mathcal{R}(\omega-\omega_{jm})
+\sin(\Theta_{jk}(\tau)\mathcal{L}(\omega-\omega_{jm})
\nonumber \\ 
+&\cos(\Theta_{jk}(\tau)\mathcal{R}(\omega+\omega_{jm})
-\sin(\Theta_{jk}(\tau)\mathcal{L}(\omega+\omega_{jm})\Big]
\label{eq:NEQ-TAS}
\end{align}

The first group of terms, weighted by the real part $\Re(P_{v,km}P_{u,mj})$, reproduces the standard result derived under the assumption of real-valued orbitals.~\cite{Walkenhorst2016} The second group, proportional to the imaginary part $\Im(P_{v,km}P_{u,mj})$, arises exclusively in theories with intrinsically complex wave functions, such as four-component Dirac theory or exact two-component Hamiltonians. This imaginary contribution vanishes identically when the system is restricted to real orbitals, marking a key distinction between nonrelativistic and relativistic formulations. Moreover, the entire nonequilibrium structure collapses to the conventional equilibrium response when the initial state is replaced by a single eigenstate, as the sums over $j$ and $k$ reduce to a single term.

The combination of Lorentzian absorptive ($\mathcal{L}$) and Rayleigh dispersive ($\mathcal{R}$) lineshapes in Eq.~\ref{eq:NEQ-TAS}, weighted by the time-dependent phase factors $\cos\theta_{jk}(\tau)$ and $\sin\theta_{jk}(\tau)$, gives rise to the characteristic signatures of transient absorption spectroscopy. Specifically, the nonequilibrium response exhibits several hallmark features absent in conventional linear absorption: (i) the $\tau$-dependent phase modulation produces spectral oscillations as a function of pump-probe delay; (ii) interference between positive- and negative-frequency components can lead to peaks with negative intensity; (iii) diagonal contributions ($j = k$) produce a $\tau$-independent background signal; and (iv) off-diagonal terms ($j \neq k$) capture the coherent evolution of the pump-generated wavepacket and allow direct observation of quantum beating phenomena.

Similarly to the TAS, we can also express the non-perturbative magnetic dipole response of the transient state $\ket{\Psi[\bm{\mathcal{E}}]}$ prepared with a pump $\bm{\mathcal{E}}$ to the probe pulse $\bm{\mathcal{F}}$ as,
$m_u[\bm{\mathcal{E}},\bm{\mathcal{F}}](t)~=~\Braket{\Psi(t)|\hat{M}_u|\Psi(t)}    
$, 
where $\hat{M}_u$ is the magnetic dipole moment operator. This response can be written in frequency domain as
\begin{align}
  \label{eq:nonpertResponse}
  m_u[\bm{\mathcal{E}},\bm{\mathcal{F}}](\omega) = i\sum_{jkmn} c^*_j(\bm{\mathcal{E}}) c_n(\bm{\mathcal{E}}) e^{i\omega_{jn}\tau} \frac{\left(e^{-i\mathcal{F}_0P_v}\right)_{jk}M_{u,km}\left(e^{i\mathcal{F}_0P_v}\right)_{mn}}{\omega_{km}+\omega+i\Gamma},
\end{align}
 The nonequilibrium linear response to the weak probe field can be obtained by expanding the exponential in Eq.~\eqref{eq:nonpertResponse} and taking the terms linear in $\mathcal{F}_0$, which gives
\begin{align}
  \label{eq:response1}  \frac{m^{(1)}_u[\bm{\mathcal{E}},\bm{\mathcal{F}}](\omega)}{\mathcal{F}_0} = \sum_{jkm} c_j(\bm{\mathcal{E}})c^*_k(\bm{\mathcal{E}}) e^{-i\omega_{jk}\tau} \frac{P_{v,km}M_{u,mj}}{\omega-\omega_{jm}+i\Gamma} + \text{c.c.}(\omega\rightarrow -\omega),
\end{align}
where c.c.$(\omega\rightarrow -\omega)$ labels the complex conjugation while
flipping the sign of the frequency.

Following the steps discussed above for TAS, we can obtain the real and imaginary parts of the electric dipole - magnetic dipole response
function in frequency domain as,~\cite{Norman2018} 
\begin{align}
  \Re\chi_{\hat{M}_u,\hat{P}_v}[\bm{\mathcal{E}}] (\omega,T+\tau) = \sum_{jkm} |&c_j(\bm{\mathcal{E}})c_k(\bm{\mathcal{E}})| \Re\left(P_{v,km}M_{u,mj}\right) \nonumber\\
  \times \big[&\cos\Theta_{jk}(\tau) \mathcal{R}(\omega-\omega_{jm}) + \sin\Theta_{jk}(\tau) \mathcal{L}(\omega-\omega_{jm}) \nonumber\\
  - &\cos\Theta_{jk}(\tau) \mathcal{R}(\omega+\omega_{jm}) + \sin\Theta_{jk}(\tau) \mathcal{L}(\omega+\omega_{jm})\big] \nonumber\\
  + \sum_{jkm} |&c_j(\bm{\mathcal{E}})c_k(\bm{\mathcal{E}})| \Im\left(P_{v,km}M_{u,mj}\right) \nonumber\\
  \times \big[&-\sin\Theta_{jk}(\tau) \mathcal{R}(\omega-\omega_{jm}) + \cos\Theta_{jk}(\tau) \mathcal{L}(\omega-\omega_{jm}) \nonumber\\
  &- \sin\Theta_{jk}(\tau) \mathcal{R}(\omega+\omega_{jm}) - \cos\Theta_{jk}(\tau) \mathcal{L}(\omega+\omega_{jm})\big],
  \label{eq:response1Real}
\end{align}
\begin{align}
  \Im\chi_{\hat{M}_u,\hat{P}_v}[\bm{\mathcal{E}}] (\omega,T+\tau) = \sum_{jkm} |&c_j(\bm{\mathcal{E}})c_k(\bm{\mathcal{E}})| \Re\left(P_{v,km}M_{u,mj}\right) \nonumber\\
  \times \big[&\sin\Theta_{jk}(\tau) \mathcal{R}(\omega-\omega_{jm}) - \cos\Theta_{jk}(\tau) \mathcal{L}(\omega-\omega_{jm}) \nonumber\\
  + &\sin\Theta_{jk}(\tau) \mathcal{R}(\omega+\omega_{jm}) + \cos\Theta_{jk}(\tau) \mathcal{L}(\omega+\omega_{jm})\big] \nonumber\\
  + \sum_{jkm} |&c_j(\bm{\mathcal{E}})c_k(\bm{\mathcal{E}})| \Im\left(P_{v,km}M_{u,mj}\right) \nonumber\\
  \times \big[ &\cos\Theta_{jk}(\tau) \mathcal{R}(\omega-\omega_{jm}) + \sin\Theta_{jk}(\tau) \mathcal{L}(\omega-\omega_{jm}) \nonumber\\
  + &\cos\Theta_{jk}(\tau) \mathcal{R}(\omega+\omega_{jm}) - \sin\Theta_{jk}(\tau) \mathcal{L}(\omega+\omega_{jm})\big].
  \label{eq:response1Imag}
\end{align}

\section{Examples}\label{sec:examples}

This section demonstrates the application of real-time time-dependent density functional theory to simulate pump-probe spectroscopies across different energy regimes and molecular systems and also to understand the underlying mechanism. Further we illustrate how non-equilibrium response theory provides physical insight into transient spectroscopic signatures.

\subsection{Pump-Probe Transient Absorption Spectroscopy (TAS)}\label{sec:TAS}

\begin{figure}[!]
    \centering
    \includegraphics[width=0.75\linewidth]{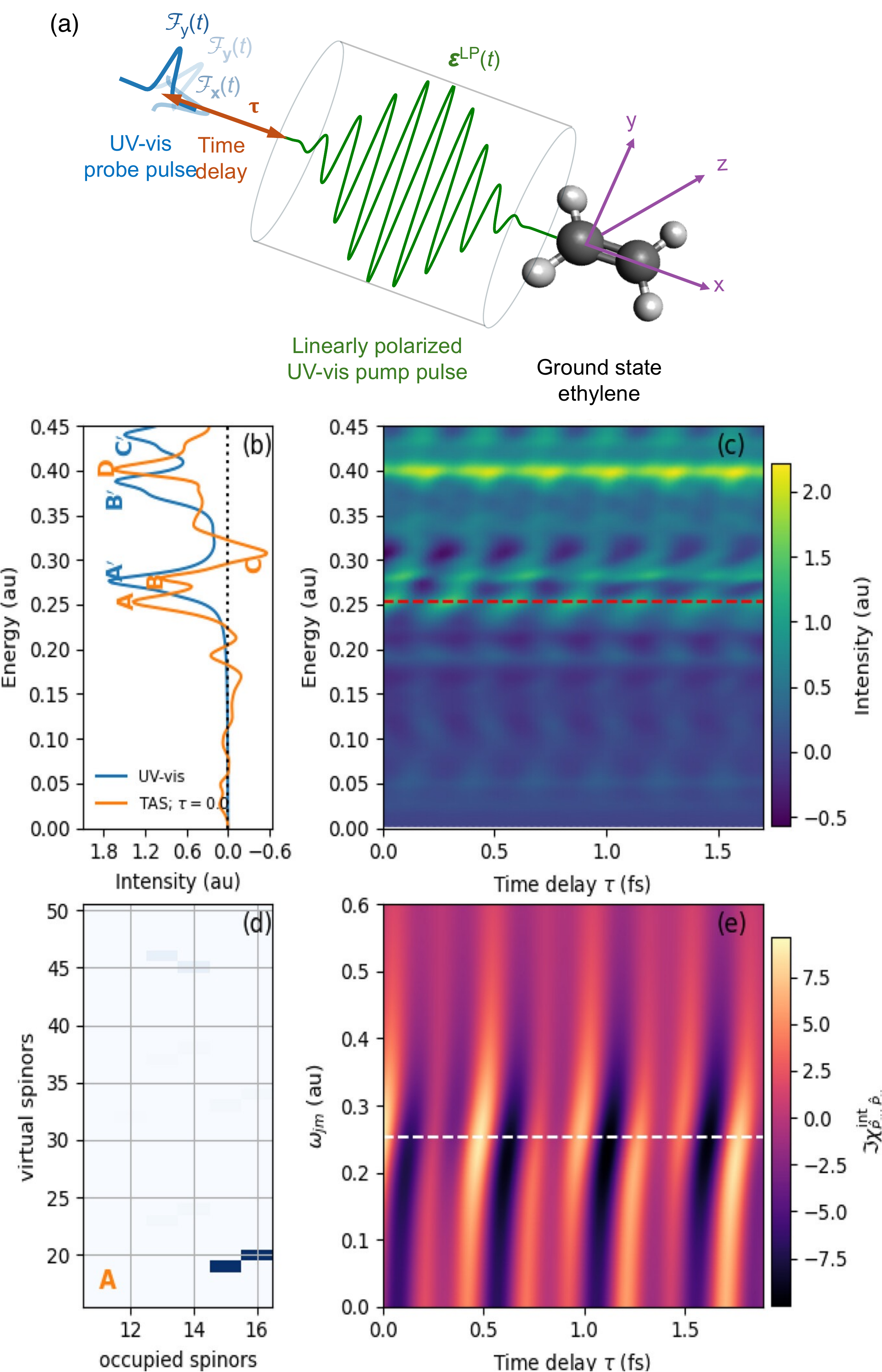}
    \caption{Ethylene: (a) Pump-probe setup; (b) ground-state absorption (blue) and transient absorption (orange) spectra at time-delay 0.0~fs; (c) TAS spectra varying with time-delay ($\tau$). 
    (d) density weighted transition analysis for peak A at $\omega'=0.25$~au;
    (e) imaginary part of electric dipole-electric dipole response function obtained from simplified non-equilibrium response function expression. 
    Adapted from Reference~\citenum{moitra2023accurate}. Copyright 2023 American Chemical Society.  }
    \label{fig:ethylene_rt}
\end{figure}

We first look into the TAS spectra of ethylene to get acquainted with the spectral imprints of the non-stationary electronic wavepacket.~\cite{moitra2023accurate} Here, a linearly polarized pump pulse (Eq.~\ref{eq:pumpLP}) tuned to the first bright excitation ($\omega_0 = 0.2762 $au), with sufficient amplitude of $\mathcal{\bm{E}}_0 = 0.01 $ au was used to substantially depopulate the ground state. A delta-type (Eq.~\ref{eq:probe}) probe pulse with $\mathcal{\bm{F}}_0= 0.01$~au is used. The setup is illustrated in Figure~\ref{fig:ethylene_rt}a.

A characteristic feature of the TAS spectra is the appearance of negative intensity peaks (see Figure~\ref{fig:ethylene_rt}b) in comparison to only positive intensity features obtained for ground stationary-state absorption. This can be attributed to the contribution of Rayleigh lineshape functions as given by Eq.~\ref{eq:NEQ-TAS}. 
The other hallmark of coherent electronic dynamics is that the spectra shows oscillatory intensity patterns with varying pump-probe time-delay.

In order to identify the underlying transitions accompanying the TAS peaks we perform dipole-weighted transition analysis (DWTA). For a resonant frequency ($\omega'$), the Fourier component $\mathcal{T}(\omega')$ of the time-domain signal $\mathcal{T}(t)$ contains the all information.~\cite{moitra2023accurate} 
\begin{align}
    \mathcal{T}(\omega') &= \int_0^\infty \mathcal{T}(t+T+\tau) e^{(i\omega'-\Gamma)t} dt
    \quad;\quad
    \mathcal{T}_{uv,ai}(t) = \Delta \mu_{uv,ai}^\mathrm{TAS}(t)
\end{align}
where, $a,i$ runs over occupied and virtual orbitals (or spinors), respectively.
DWTA analysis revealed that peaks A, B, and C arise dominantly from two degenerate spinor-pair excitations: $(15,16) \to (19,20)$ and $(13,14) \to (46,45)$, 
as shown in Figure~\ref{fig:ethylene_rt}d. Now, using a simplified version of Eq.~\ref{eq:NEQ-TAS}, assuming that (i) only these two transitions contribute to the TAS, (ii) have equal transition probability, (iii) have zero phase-factor and (iv) have the uniform electric dipole transition moments, one can still capture the essential TAS spectral beating pattern as shown in Figure~\ref{fig:ethylene_rt}e.
Note that this oscillation of spectral intensity with $\tau$ does not arise from changes in the contributing orbitals but from coherent interference effects encoded in the non-equilibrium response function.

\begin{figure}[htb!]
    \centering
    \includegraphics[width=\linewidth]{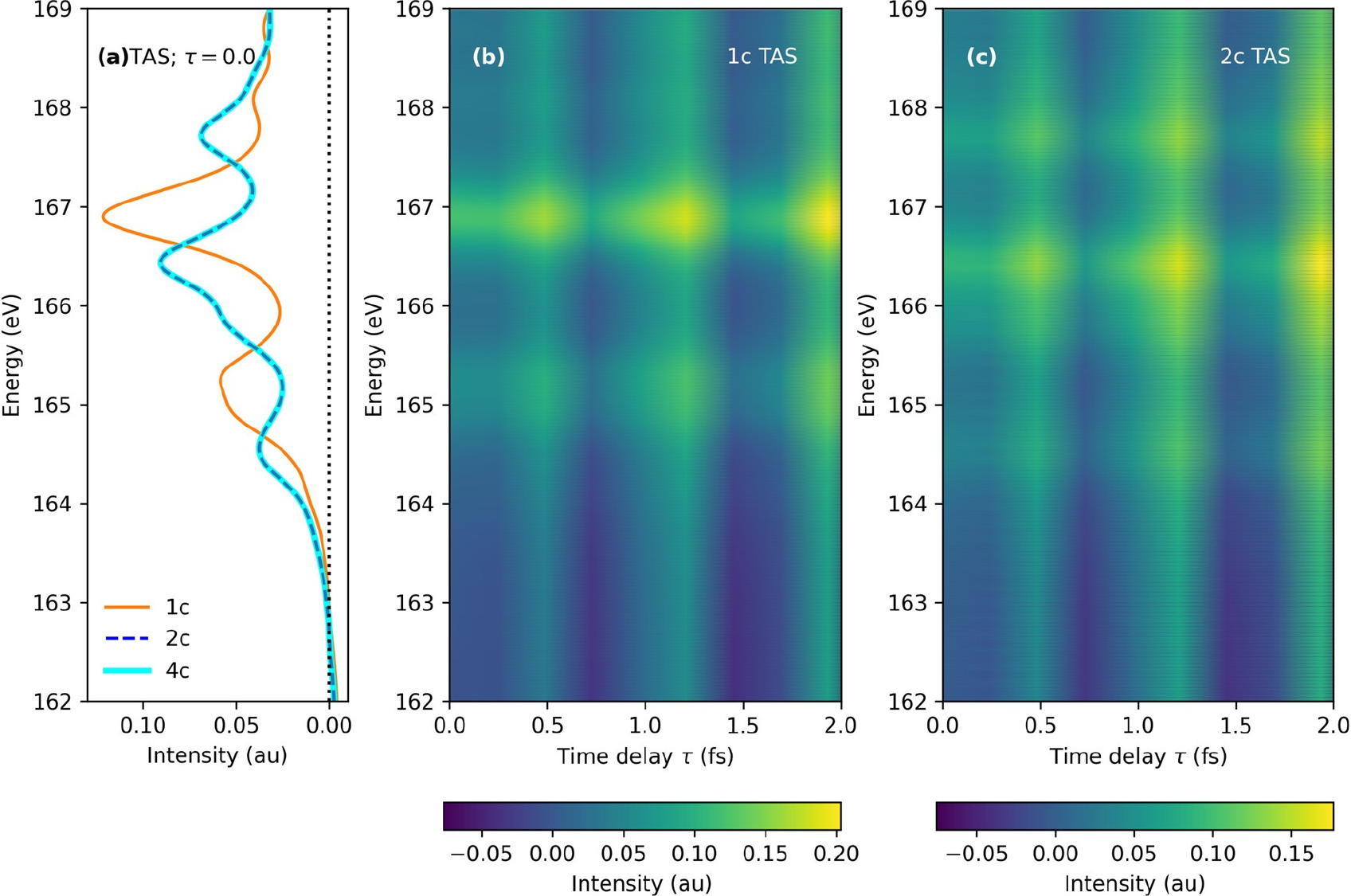}
    \caption{Thiophene: (a) Sulphur L$_{2,3}$-edge x-ray TAS at time-delay $\tau=0.0$~fs computed using 1c non-relativistic (orange), 2c amfX2c (blue) and 4c Dirac--Coulomb (cyan) Hamiltonians. 
    Variation of TAS spectra with pump-probe time-delay computed at (b) 1c non-relativistic and (c) 2c amfX2C Hamiltonian levels. Reproduced from Reference~\citenum{moitra2023accurate}. Copyright 2023 American Chemical Society.}
    \label{fig:thiophene}
\end{figure}

Next, in order to demonstrate the importance of relativistic effects, especially when the frequency region of interest lies in the X-ray regime, we use the example of thiophene, as shown in Figure~\ref{fig:thiophene}.~\cite{moitra2023accurate} 
In accordance with the previous example, the carrier frequency of the pump pulse was set to the first excitation energy ($\omega_0 = 0.2135$~ au). After the pump excitation, X-ray transient absorption of the wavepacket at the sulfur L$_{2,3}$-edges was probed. The spectrum clearly shows the spin-orbit splitting of the sulfur 2p orbitals. This feature can only be described correctly using relativistic methods.
The key observations from the simulation results presented in Figure~\ref{fig:thiophene} are that:
(i) relativistic Hamiltonian description is needed to adequately describe the scalar as well as spin-orbit relativistic imprints. 
(ii) the 2c amfX2C method provides a reliable, accurate yet computationally cheaper framework to get reference 4c Dirac-Coulomb Hamiltonian quality spectral results. 
(iii) The TAS spectra varies with pump-probe delay showing alternating low- and high-intensity regions which marks the time evolution of the superposition state created by the pump pulse.

\subsection{Pump-Probe Transient Electronic Circular Dichroism}\label{sec:TCD}

Conventional ground-state electronic circular dichroism (ECD) spectroscopy measures the differential absorption of left- and right-circularly polarized light and is used to characterize structurally chiral molecules. Recent studies have shown that circularly polarized laser pulses can induce purely electronic chirality in achiral molecules without altering nuclear geometry, a phenomenon termed as {light-induced electronic chirality}.~\cite{Moitra2025,Chen2024} Further to detect these chiral wavepackets, transient electronic circular dichroism spectroscopy was theoretically proposed.~\cite{Moitra2025} 

\begin{figure}[htb!]
    \centering
    \includegraphics[width=1.0\linewidth]{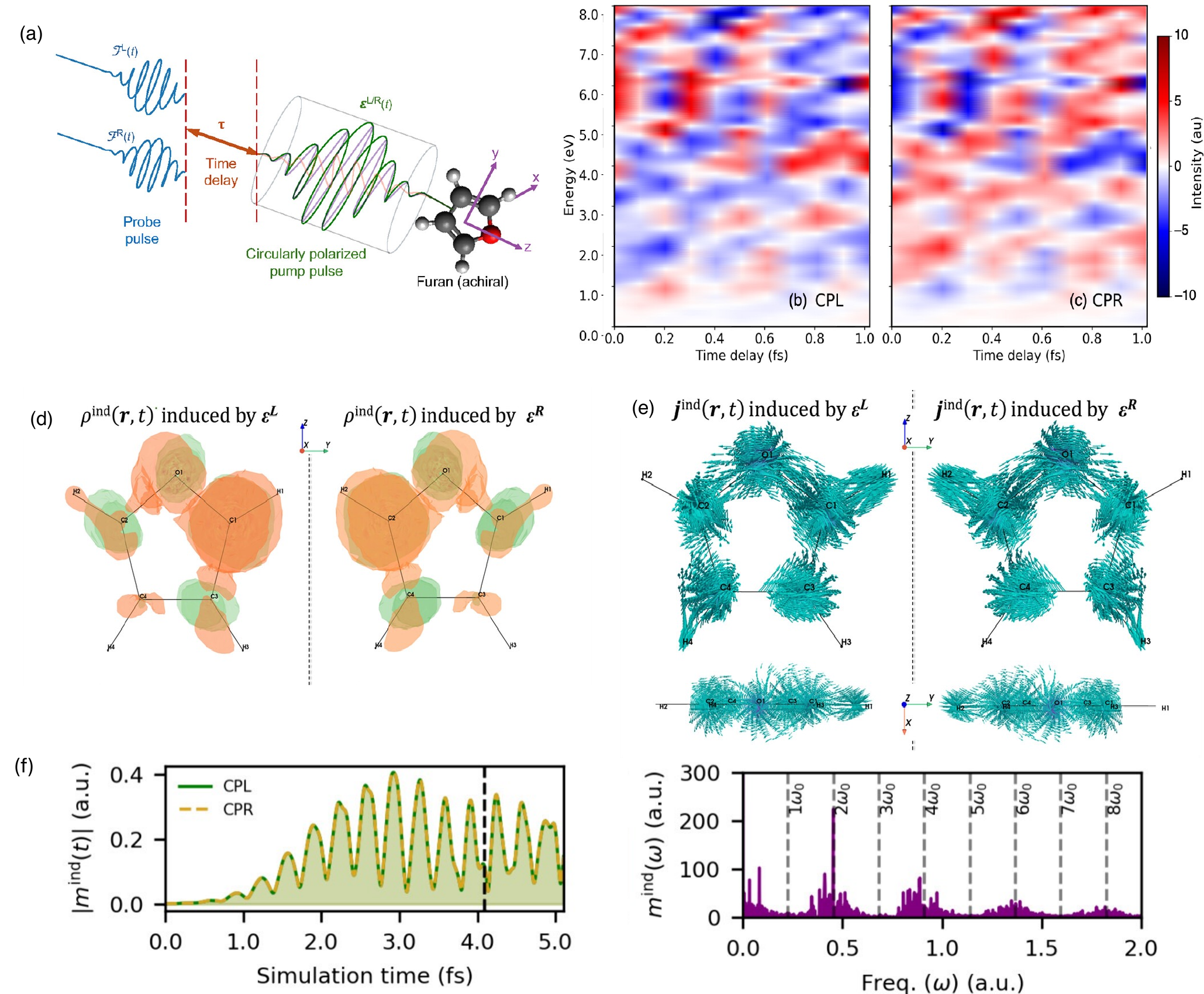}
    \caption{Furan: (a) Pump-probe setup; time-delay dependent transient electronic circular dichroism spectra of aligned furan pumped with (b) left and (c) right circularly polarized light;
    chiral induced (d) electron charge density $\rho^\mathrm{ind}(\bm{r}, t)$ and (e) current density $\vect{j}^{\mathrm{ind}}(\vect{r}, t)$ generated by the circularly polarized light at the end of the pump pulse (t = T); (f) The time-dependent induced magnetic dipole moment and it´s Fourier transform. Adapted from Reference~\citenum{Moitra2025}. Copyright 2025 American Chemical Society.}
    \label{fig:furan}
\end{figure}

This effect was demonstrated using oriented furan molecule. The computational setup is shown in Figure~\ref{fig:furan}a, where an achiral furan molecule is pumped by a circularly polarized (CP) laser pulse with direction of propagation aligned with the static electric dipole moment of the system. An isotropically averaged delta probe pulse is applied after a time-delay which mimics the differential absorption of left and right circularly polarized light. The TCD spectra obtained by left and right circularly polarized pump light (Figure~\ref{fig:furan}b,c) shows mirror-image symmetry. This feature is indicative of enantiomeric relationship of the induced chiral wavepacket. The same can be observed, as the charge and current density induced by the circularly polarized left (CPL) and right (CPR) pulse computed at the end of the pump pulse possesses non-superimposable mirror-image symmetry, as shown in Figure~\ref{fig:furan}d,e. This mirror-image relationship persists beyond the end of the pump pulse. 
However, it is also observed that there is no single timescale at which the spectra flips sign, or one at which the electronic wavepacket flips chirality. Instead different spectral energy regions exhibit different time-periods of change in sign of spectral signal. This observation is linked to the multiple harmonic order of the carrier frequency that contribute to the Fourier transform of the induced magnetic dipole moment, as shown in Figure~\ref{fig:furan}f.

\begin{figure}[htb!]
    \centering
    \includegraphics[width=1.0\linewidth]{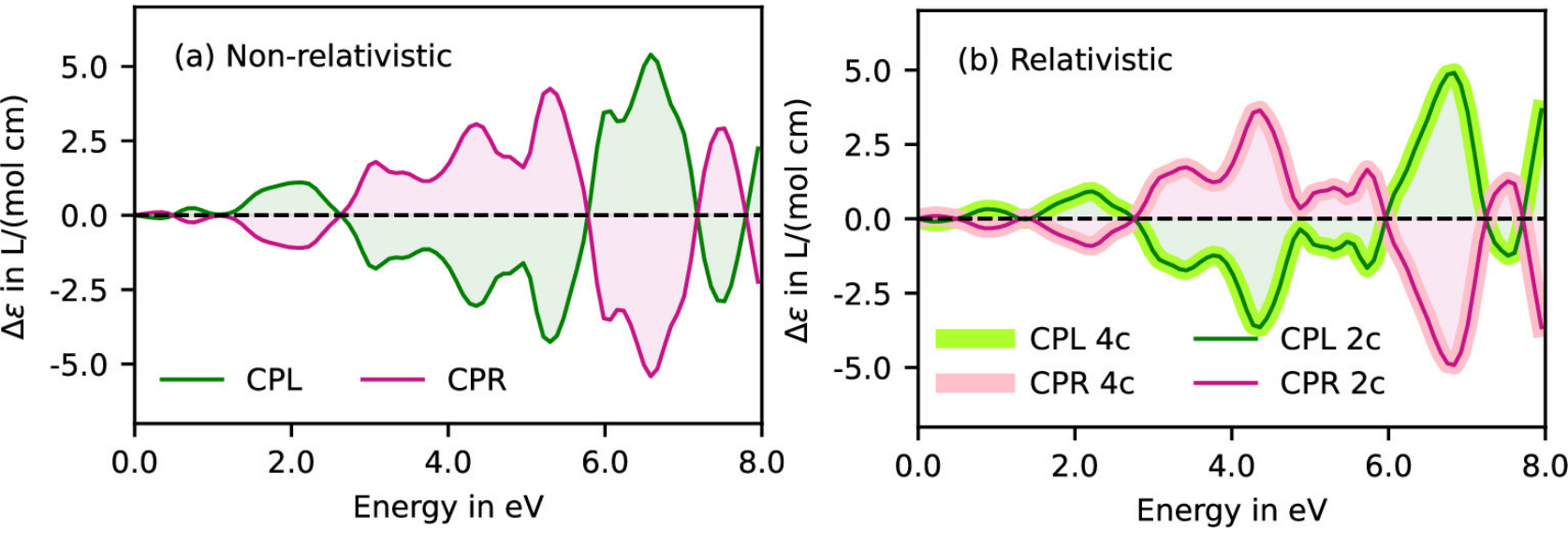}
    \caption{Tellurophene: Transient electronic circular dichroism spectra of alligned molecule at time-delay $\tau=0.0$~fs with circularly polarized left and right pump pulses obtained with (a) 1c non-relativistic and (b) relativistic Hamiltonians, namely 2c amfX2C and 4c Dirac--Coulomb. Reproduced from Reference~\citenum{Respect2025}. Copyright 2025 American Chemical Society. }
    \label{fig:tellurophene}
\end{figure}

The extent of relativistic effect on light-induced electronic chirality can further be estimated by looking at the TCD spectra obtained by replacing the furan molecule by its Tellurium analogue (tellurophene).~\cite{Respect2025} The computational setup is consistent with the one presented in Figure~\ref{fig:furan}a. The spectral functions in Figure~\ref{fig:tellurophene} are reported for valence energy regions, but the comparison underscores that accurate treatment of relativistic effects is essential for quantitative predictions of TCD in molecules containing fourth-row and heavier elements, where SO coupling can dramatically reshape the spectroscopic response.
Consistent with previous observations from Figure~\ref{fig:thiophene}a, we see a remarkable agreement between the 2c amfX2C and 4c Dirac--Coulomb methodology. This further enhances the confidence on the amfX2C methodology to be used regularly for studying ultrafast pump-probe processes. 

These results on furan and tellurophene demonstrate that a single monochromatic CP pulse can induce long-lived electronic chirality detectable via attosecond transient absorption spectroscopy, opening pathways for ultrafast chiral state control in spintronics and enantioselective chemistry.

\section{Conclusions}\label{sec:summary}

This chapter establishes real-time TDDFT as a versatile and computationally tractable framework for the simulation of pump-probe spectroscopies across spectral regimes. The non-equilibrium response theory provides a transparent analytical account of the spectral features that distinguish transient spectra from conventional ones, identifying the roles of pump-prepared wavepacket, coherence etc. The application cases demonstrate the need for different Hamiltonians tailored to the problem, for instance, relativistic treatment is necessary for core-level spectroscopies and systems containing heavy elements. Furthermore, the case of light-induced electronic chirality shows the potential of theoretical approaches to guide experiments in new directions. 

However, several methodological limitations of the presented framework exist.
First, the simulations described here are based on localized atomic orbital basis sets, widely used in quantum chemistry for bound states. This choice is pragmatic because it allows for easy integration into existing quantum chemistry packages and efficient evaluation of one- and two-electron integrals. However, it has a fundamental drawback for processes involving the ionization of electrons into the continuum, such as in strong-field ionization. A standard strategy to overcome this is to include a complex absorbing potential at the periphery of the molecular system that absorbs outgoing electron density or use specialized continuum basis functions.~\cite{Riss1996CAP,Vibok1992CAP,Zhu2022CAP,bspline2020,pade2021,bspline2022cc3,bspline2021jctc} 
Second, the closed-system Liouville-von Neumann equation discussed here, conserves the purity of the density matrix, while real pump--probe experiments are effected by decoherence originating from vibrational motions, coupling to solvent or bath. A rigorous treatment of these effects would require embedding the RT-TDDFT electron propagation within an open quantum system framework, where the electronic subsystem is coupled to a bath.~\cite{Tempel2011Lindblad,Tempel2011LRtddft,Floss2019openQS,Chen2018OpenQs}
Additionally, the frozen-nuclei approximation is physically meaningful only upto the first few-femtoseconds; extending the framework to coupled nuclear-electronic dynamics through mean-field Ehrenfest approach or more advanced surface hopping or MCTDH will be necessary for simulations at longer delays.~\cite{li2005mmut,Li2025neo,Zhao2020neoEhrenfest,Meyer1990} Finally, the adiabatic approximation for the exchange-correlation potential neglects memory effects and hence all the precautionary measures known for conventional linear-response TDDFT extends to RT-TDDFT.~\cite{Schirmer2025tddft,Ullrich2025TDDFT,Dreuw2004TDDFT}

\begin{acknowledgement}

\end{acknowledgement}
T.M. thanks Dr. Michal Repisky for fruitful discussions and the Research Initiation Grant from the Indian Institute of Technology Bhilai for financial support. 
\bibliography{references}

\end{document}